\documentstyle[12pt]{article}
\def\l#1#2{\raisebox{.2ex}{$\displaystyle
  \mathop{#1}^{{\scriptstyle #2}\rightarrow}$}}
\def\r#1#2{\raisebox{.2ex}{$\displaystyle
 \mathop{#1}^{\leftarrow {\scriptstyle #2}}$}}

\makeatletter
\def\eqnarray{\stepcounter{equation}\let\@currentlabel=\theequation
\global\@eqnswtrue
\global\@eqcnt\z@\tabskip\@centering\let\\=\@eqncr
$$\halign to \displaywidth\bgroup\@eqnsel\hskip\@centering
  $\displaystyle\tabskip\z@{##}$&\global\@eqcnt\@ne
  \hfil$\displaystyle{\hbox{}##\hbox{}}$\hfil
  &\global\@eqcnt\tw@ $\displaystyle\tabskip\z@
  {##}$\hfil\tabskip\@centering&\llap{##}\tabskip\z@\cr}
\@addtoreset{equation}{section}
  \def\theequation{\thesection.\arabic{equation}}
\makeatother

\begin{document}

\renewcommand{\thefootnote}{\fnsymbol{footnote}}
\newpage
\setcounter{page}{0}
\pagestyle{empty}
\begin{flushright}
{November 1996}\\
{JINR E2-96-410}\\
{hep-th/9611108}
\end{flushright}
\vfill

\begin{center}
{\LARGE {\bf  $N=2$ superintegrable f-Toda mapping }}\\[0.3cm]
{\LARGE {\bf  and super-NLS hierarchy  }}\\[0.3cm]
{\LARGE {\bf  in $(1|2)$ superspace }}\\[1cm]

{\large V.B. Derjagin$^{a,1}$, A.N. Leznov$^{a,2}$ and A.S. Sorin$^{b,3}$}
{}~\\
\quad \\
{\em {~$~^{(a)}$ Institute for High Energy Physics,}}\\
{\em 142284 Protvino, Moscow Region, Russia}\\
{\em {~$~^{(b)}$ Bogoliubov Laboratory of Theoretical Physics, JINR,}}\\
{\em 141980 Dubna, Moscow Region, Russia}~\quad\\

\end{center}

\vfill

\centerline{ {\bf Abstract}}
A new integrable $N=2$ supersymmetric f-Toda mapping in $(1|2)$
superspace, acting like the symmetry transformation of $N=2$
supersymmetric NLS hierarchy, is proposed. The first two
Hamiltonian structures and the recursion operator connecting all evolution
systems and Hamiltonian structures of the $N=2$ super-NLS hierarchy are
constructed in explicit form using only invariance conditions
with respect to the f-Toda mapping. A new representation for its
Hamiltonians is observed.

\vfill
{\em E-Mail:\\
2) derjagin@mx.ihep.su\\
2) leznov@mx.ihep.su\\
3) sorin@thsun1.jinr.dubna.su }
\newpage
\pagestyle{plain}
\renewcommand{\thefootnote}{\arabic{footnote}}
\setcounter{footnote}{0}

\section{Introduction}

Recently, a manifestly $N=2$ supersymmetric formulation of the Nonlinear
Schr\"odinger (NLS) hierarchy (see, e.g., \cite{1} and references therein)
has been constructed \cite{2,3}. The method is based
on the super-Hamiltonian formalism, sufficiently cumbersome for
some concrete calculations. Quite recently, the advantage of an
alternative approach to the theory of supersymmetric integrable
hierarchies has been demonstrated in \cite{LS}, where the formalism of
integrable mappings \cite{5,6} was applied to the problem of constructing
hierarchies of $(1+2)$-dimensional integrable systems in $(2|2)$
superspace. In \cite{LS}, a few two-dimensional superintegrable mappings
were proposed. It would be interesting to find some new examples
manifesting the benefit of using supersymmetric mappings.

The goal of the present Letter is to present the results of Refs.
\cite{1,2,3} (including the derivation of some new results) as a direct
corollary to the existence of a new integrable $N=2$ supersymmetric
mapping (we also call it a substitution) acting in $(1|2)$
superspace. This mapping relates two pairs of chiral-antichiral fermionic
superfields and, in the bosonic limit, it is equivalent to
the one-dimensional Toda mapping. Taking into account that the
Toda-mapping is responsible for the existence and properties of the
bosonic NLS hierarchy, we call our supersymmetric mapping Fermi-Toda
(f-Toda), reflecting the existence of the fermionic fields in its
background. However, this name may be considered to have a deeper
foundation if one remembers that Fermi was one of the authors of
\cite{4}, where equations of a nonlinear chain were applied for the first
time to the solution of the physical problem of establishing the heat
equilibrium in a short-range interacting dynamic system.

\section{$N=2$ supersymmetric f-Toda mapping}

In this section, we introduce the f-Toda mapping and show its
integrability.

We work in $(1|2)$ superspace with one bosonic $x$ and two fermionic
$\theta,\overline \theta$ coordinates and use standard representation
for the $N=2$ supersymmetric fermionic covariant derivatives
\begin{eqnarray}
D=\frac{\partial}{\partial \theta}-{1\over 2}\overline \theta \frac{\partial}
{\partial x},\quad \overline D=\frac{\partial}{\partial \overline
\theta}-{1\over 2}\theta\frac{\partial}{\partial x}, \quad
D^2= (\overline D)^2=0, \quad \{D,\overline D \}=
-\frac{\partial}{\partial x} \equiv -{\partial}.
\label{1}
\end{eqnarray}

Let us introduce a pair of chiral and antichiral fermionic
superfields $f(x,\theta,\overline \theta)$ and
${\overline f}(x,\theta,\overline \theta)$, respectively,
\begin{equation}
Df={\overline D}~{\overline f}=0,
\label{2}
\end{equation}
and the following relation\footnote{The sign $'$ means the
derivative with respect to $x$.}:
\begin{equation}
{1\over 2}(\r {f}{} \r {\overline f}{}-f \overline f)=
(\ln (\overline D \r {f}{}\cdot D \overline f))',
\label{3}
\end{equation}
which is the definition of the mapping or the rule determining the
correspondence between two initial functions, $f$ and $\overline f$, and two
final ones, $\r {f}{}$ and $\r {\overline f}{}$. The action of the inverse
transformation is denoted by $\l {f}{}$ and $\l {\overline f}{}$,
and the corresponding mapping takes the form
\begin{equation}
{1\over 2}(f\overline f-\l {f}{} \l {\overline f}{})=
(\ln ( \overline D f\cdot D \l {\overline f}{}))'.
\label{4}
\end{equation}
The notation $\r {f}{}$ ($\l {f}{}$) means that the index of variable $f$
is shifted by $+1$ ($-1$) (e.g., $\r {f}{}_n=f_{n+1}$).
Relations (\ref{3}) and (\ref{4}) fix only
the scaling dimension of the product $[f{\overline f}]=cm^{-1}$.

We would like to remark that the mapping (\ref{3}) is not algebraically
solvable with respect to the superfields $f$ and $\overline f$, or $\r
{f}{}$ and $\r {\overline f}{}$, in contrast to its bosonic
limit\footnote{Let us remember that in the bosonic limit, all fermionic
components must be set equal to zero.}---the Toda chain---where the
bosonic components of the superfields $f$ and $\overline f$ can be
expressed pure algebraically in terms of their counterparts belonging to
the superfields $\r {f}{}$ and $\r {\overline f}{}$, and vice versa.

Substitution (\ref{3}) possesses the inner automorphism
$\sigma$ with the properties
\begin{eqnarray}
\sigma f {\sigma}^{-1}=\r {\overline f}{}, \quad
&& \sigma {\overline f} {\sigma}^{-1}=\r f{}, \quad
\sigma \r {\overline f}{} {\sigma}^{-1}=f, \quad
\sigma \r f{} {\sigma}^{-1}={\overline f}, \nonumber\\
&& \sigma D{\sigma}^{-1}={\overline D} , \quad
\sigma {\overline D}{\sigma}^{-1}=D,
\label{auto}
\end{eqnarray}
which will be useful in what follows.
The action of $\sigma$ on the covariant derivatives $D$ and ${\overline D}$
can be induced by the following transformation of the $(1|2)$ superspace
coordinates
\begin{eqnarray}
\sigma x {\sigma}^{-1}=x, \quad
\sigma {\theta} {\sigma}^{-1}={\overline {\theta}}, \quad
\sigma {\overline {\theta}} {\sigma}^{-1}={\theta}.
\label{auto1}
\end{eqnarray}

To establish the connection of the mapping (\ref{3}) with the theory of
integrable systems, one can consider the general representation for an
evolution-type system
\begin{equation}
\frac{\partial f}{\partial t}=
F(f,\overline f, f~',{\overline f}~', {\overline D}f, D{\overline f}...),
\quad
\frac{\partial {\overline f}}{\partial t}=
{\overline F}(f,\overline f, f~',{\overline f}~',
{\overline D}f, D{\overline f}...),
\label{6}
\end{equation}
with the additional requirement of its
invariance with respect to the mapping. The system (\ref{6})
is invariant with respect to the transformation (\ref{3}) if the
functions $F$ and ${\overline F}$ are subjected to a set of constraints
called the symmetry equations of the mapping \cite{5,6}. In other words,
the symmetry equation may be treated as the condition ensuring the
invariance of the evolution-type system (\ref{6}) with respect to the
mapping (\ref{3}). It appears that it strictly determines some class of
partial solutions corresponding to some hierarchy of integrable systems.
In what follows, we show that the symmetry equation of the mapping
(\ref{3}) does extract the $N=2$ super-NLS hierarchy of integrable
equations \cite{2,3}.

The symmetry equation for a given mapping can be obtained by taking
its derivative with respect to an arbitrary parameter and denoting the
derivatives of the independent functions involved in a substitution by
correspondingly new symbols \cite{5,6}. In the case under consideration,
these symbols are $F=\dot f$ and $\overline F=\dot {\overline f}$, which
are the chiral and antichiral superfields, respectively,
\begin{equation}
DF={\overline D}~{\overline F}=0,
\label{ch1} \end{equation}
as $f$ and $\overline f$
(\ref{2}). Using this method, one can obtain a symmetry equation
corresponding to the substitution (\ref{3}) in the following form:
\begin{equation}
{1\over 2}(\r {F}{} \r {\overline f}{}+\r {f}{} \r {\overline F}{}-
F \overline f-f\overline F)=
({{\overline D} \r {F}{} \over {{\overline D} \r {f}{}}} +
{D {\overline F} \over D {\overline f}})'.
\label{5}
\end{equation}

It is obvious that (\ref{5}) possesses the trivial partial
solution $F=f',\overline F=\overline f'$. To understand this,
it is sufficient to choose the bosonic coordinate $x$ as the parameter of
differentiation. The next obvious but nontrivial solution is
\begin{equation}
F=f,\quad \overline F=-\overline f.
\label{sol}
\end{equation}
In Refs. \cite{5,6}, a mapping was called integrable if
its symmetry equation possessed at least one nontrivial solution.
Thus, in this sense, the f-Toda substitution (\ref{3}) is integrable.

To conclude this section, we state the following proposition.

{\it The f-Toda mapping (\ref{3}) is integrable and each solution of
its symmetry equation (\ref{5}) is connected with an evolution-type system
(\ref{6}) invariant with respect to its transformation}.

\section{The symmetry of the symmetry equation}

In this section, we construct a recurrent procedure for finding an
infinite set of partial solutions to the symmetry equation (\ref{5})
and establish their connection to the $N=2$ super-NLS hierarchy.

Let us present the explicit form of the transformation that generates a
new solution to the symmetry equation from an arbitrarily
given one. We make the following assertion.

{\it If the pair $F$ and $\overline F$ is a solution of the symmetry
equation (\ref{5}), the pair $\widetilde F$ and $\widetilde
{\overline F}$, defined as\footnote{Here, the derivatives $\partial$,
${\overline D}$ and $D$ act like operators, i.e.,
they must be commuted with $f$ and ${\overline f}$.}
\begin{eqnarray}
\widetilde F &=& F'+ D(f\overline D + {1\over 2}\overline Df)\partial^{-1}
(f\overline F+ F\overline f), \nonumber\\
\widetilde {\overline F}&=&-{\overline F}'+
\overline D({\overline f}D +
{1\over 2}D{\overline f})\partial^{-1}(f\overline F+F\overline f),
\label{7}
\end{eqnarray}
is also a solution}.

One can prove (\ref{7}) by straightforward but rather
tedious calculations. The main steps to prove this statement are given in
the appendix.

Representing (\ref{7}) in the form
\begin{equation}
\left(\begin{array}{cc}
\widetilde F \\ {\widetilde {\overline F}}
\end{array}\right) = R\left(\begin{array}{cc}
F\\{\overline F}
\end{array}\right),
\label{set0}
\end{equation}
one can obtain the following expression for the recursion operator $R$
of the integrable hierarchy corresponding to the substitution
(\ref{3})$^{3}$:
\begin{eqnarray}
R=\Pi \left(\begin{array}{cc}
{\partial}+ fD\overline D \partial^{-1} \overline f+
{1\over 2}{\partial} f\partial^{-1}\overline f,
& - fD\overline D \partial^{-1} f-
{1\over 2}{\partial} f{\partial}^{-1} f\\
{\overline f}~{\overline D} D \partial^{-1} \overline f+{1\over 2}
{\partial}\overline f\partial^{-1}\overline f, &
-{\partial}-{\overline f}~{\overline D} D \partial^{-1} f-
{1\over 2}{\partial} \overline f \partial^{-1} f\end{array}\right) \Pi,
\label{recop}
\end{eqnarray}
where $\Pi$ (${\overline {\Pi}}$)
\begin{eqnarray}
\Pi & \equiv & - \left(\begin{array}{cc} D\overline D \partial^{-1}, & 0 \\
0, & \overline D D \partial^{-1}
\end{array}\right), \quad
{\overline {\Pi}} \equiv - \left(\begin{array}{cc} {\overline D}
D \partial^{-1}, & 0 \\
0, & D{\overline D}  \partial^{-1}
\end{array}\right), \nonumber\\
&& \Pi \Pi =\Pi, \quad {\overline \Pi}~ {\overline \Pi}=\overline \Pi,
\quad \Pi \overline \Pi=\overline \Pi \Pi=0, \quad \Pi + \overline \Pi =1
\label{pi}
\end{eqnarray}
is the matrix that projects the up and down elements of a column on
the chiral (antichiral) and antichiral (chiral) subspaces, respectively.
Let us stress that the expression for $R$ is defined up to an arbitrary
additive operator which annihilates the column on the r.h.s. of
relation (\ref{set0}). It is clear that such an operator can be
represented in the following general form: $C{\overline {\Pi}}$, where $C$
is an arbitrary matrix-valued pseudo-differential operator.

Simple inspection of $R$ (\ref{recop}) shows that it possesses the
following properties:
\begin{eqnarray}
\Pi R=R \Pi=R, \quad {\overline {\Pi}}R=R{\overline {\Pi}} =0,
\label{prop}
\end{eqnarray}
and, therefore, its action preserves the chiral structure (\ref{ch1}) of
the evolution equations (\ref{6}). Because of this chiral structure, all
expressions for the recursion operator, which differ by the above-mentioned
operator $C {\overline {\Pi}}$, are equivalent.

Acting $p$-times ($p=0,1,2, ...$) by the recursion
operator on the first nontrivial solution (\ref{sol}) of the symmetry
equation, we can generate the new solutions $F_p$ and ${\overline F}_p$,
\begin{equation}
\left(\begin{array}{cc}
F_p\\{\overline F}_p
\end{array}\right) = R~^{p}
\left(\begin{array}{cc}
f\\-{\overline f}
\end{array}\right),
\label{set}
\end{equation}
and the corresponding evolution equations
\begin{equation}
\frac{\partial}{\partial t_p}
\left(\begin{array}{cc} f \\ {\overline f} \end{array}\right)=
\left(\begin{array}{cc} F_p\\{\overline F}_p \end{array}\right)
\label{evol}
\end{equation}
belonging to the integrable hierarchy. Taking into account the scaling
dimension of the recursion operator $[R]=cm^{-1}$, it is easy to observe
that for the $p$-th solution, the maximal order of the bosonic derivative
linearly appearing on the right-hand side of eqs. (\ref{evol}) is
equal to $p$. Using the terminology of inverse scattering theory,
one can say that the $p$-th solution corresponds to the $p$-th flow.

The first five solutions to the symmetry equation have the following form:
\begin{eqnarray}
&&F_0 = f \; , \quad {\overline F}_0 = -{\overline f}; \; \quad
F_1 = f' \; , \quad {\overline F}_1 = {\overline f}'; \; \nonumber\\
&&F_2 =  f'' + D(f{\overline f}~{\overline D} f), \quad
{\overline F}_2 =-{\overline f}'' +
{\overline D}(f{\overline f} D{\overline f}); \; \nonumber\\
&&F_3 = f''' + \frac{3}{2} D((f{\overline D}f)'\overline f), \quad
{\overline F}_3 = {\overline f}''' +
\frac{3}{2} {\overline D}(({\overline f}D{\overline f})'f) ;
\nonumber\\
&&F_4=f''''+\frac{1}{2}D[3(f{\overline D}f)''{\overline f}-
3f{\overline f} D{\overline f} \cdot ({\overline D}f)^2+
f({\overline f}~{\overline D}f)''+
(f{\overline f})''{\overline D}f], \nonumber\\
&&{\overline F}_4=-{\overline f}''''-
\frac{1}{2}{\overline D}[3({\overline f}D{\overline f})''f+
3f{\overline f}~{\overline D}f\cdot(D{\overline f})^2+
{\overline f}(fD{\overline f})''-
(f{\overline f})''D{\overline f}].
\label{three}
\end{eqnarray}
These expressions coincide with the corresponding ones
for the $N=2$ supersymmetric NLS hierarchy \cite{2,3}, therefore, we can
recognize that the f-Toda mapping (\ref{3}) is related to the $N=2$
supersymmetric NLS hierarchy, which justifies its name.

Thus, the following proposition summarizes this section.

{\it The f-Toda mapping (\ref{3}) acts like the symmetry transformation of
the $N=2$ super-NLS hierarchy}.

\section{The f-Toda invariant Hamiltonian structures}

In this section, we construct the Hamiltonian structures which are
invariant with respect to the f-Toda mapping (\ref{3}).

By definition, for the chiral-antichiral fermionic superfields $f$ and
${\overline f}$, the f-Toda invariant Hamiltonian structure $J$ is
a symmetric\footnote{Let us recall the rules of the adjoint conjugation
operation ``$T$'': $D^{T}=-D$, ${\overline D}^{T}=-{\overline D}$,
$(MN)^{T}=(-1)^{d_Md_N}N^{T}M^{T}$, where $d_M$ ($d_N$) is the Grassman
parity of the operator $M$ ($N$), equal to 0 (1) for bosonic (fermionic)
operators. In addition, for matrices, it is necessary to take the operation
of the matrix transposition. All other rules can be derived using
these.} $J^{T}=J$ pseudo-differential $2$x$2$ matrix operator which,
in addition to the Jacobi identity and the chiral consistency conditions
\begin{eqnarray}
J \Pi={\overline \Pi} J=0, \quad J {\overline \Pi}= \Pi J=J,
\label{cons}
\end{eqnarray}
should also satisfy the following additional constraint \cite{7,8}:
\begin{eqnarray}
J(f, {\overline f})={\Phi}J(\r {f}{}, \r {\overline f}{}){\Phi}^T,
\label{inv2}
\end{eqnarray}
which provides its invariance with respect to the f-Toda mapping.
Here$^{3}$,
\begin{eqnarray}
{\Phi}={\phi}_1\otimes {\phi}_2\equiv
\Pi \left(\begin{array}{cc}
D(1-\frac{1}{2}fL^{-1}{\overline f}) \\
-\frac{1}{2}\{D,{\overline f}\}L^{-1} {\overline f}
\end{array}\right)\{D,{\overline f}\}^{-1} \otimes
\left(\begin{array}{cc}
\r {\overline f}{}+2{\overline D}{\partial}\{{\overline D},\r f{}\}^{-1},
& - \r  f{} \end{array}\right) \Pi
\label{frech}
\end{eqnarray}
is the inverse matrix of Fr\'echet derivatives corresponding to the
mapping (\ref{3}), where the notation `$\otimes$' stands for the tensor
product and the operator $L$,
\begin{eqnarray} L \equiv {\partial} -\frac{1}{2}f{\overline f} -
\frac{1}{2}f{\partial}^{-1}{\overline D}\{D,{\overline f}\},
\quad [D,L]=0
\label{lax}
\end{eqnarray}
coincides with the Lax operator of the $N=2$ supersymmetric NLS hierarchy
\cite{3}. One can easily invert relation (\ref{inv2}) by applying the
automorphism $\sigma$ (\ref{auto}). As a result, we obtain
\begin{eqnarray}
J(\r {f}{}, \r {\overline f}{})=
{\hat{\Phi}} J(f,{\overline f}){\hat{\Phi}}^T,
\label{inv4}
\end{eqnarray}
where
\begin{equation}
{\hat{\Phi}}=A\sigma {\Phi}{\sigma}^{-1}A \equiv
{\hat{\phi}}_2\otimes {\hat{\phi}}_1\equiv
(-A\sigma{\phi}_1 {\sigma}^{-1}) \otimes
(-\sigma {\phi}_2 {\sigma}^{-1}A),
\label{m}
\end{equation}
and
\begin{equation}
A \equiv \left(\begin{array}{cc} 0, & 1 \\1, & 0 \end{array}\right)
\label{a}
\end{equation}
is the matrix of Fr\'echet derivatives for the automorphism $\sigma$
(\ref{auto}). The matrices
${\hat{\phi}}_{1,2}$, ${\phi}_{1,2}$, $\Phi$, and ${\hat\Phi}$ possess the
following useful properties:
\begin{eqnarray}
&& {\hat{\phi}}_1 {\phi}_1={\phi}_2{\hat{\phi}}_2=1,
\quad {\phi}_1 \otimes {\hat{\phi}}_1 =
{\hat{\phi}}_2\otimes{\phi}_2  =
\Phi{\hat\Phi} = {\hat\Phi} \Phi = \Pi, \nonumber\\
&& \Phi \r R{}=R\Phi, \quad {\hat\Phi} R=\r R{} {\hat\Phi}, \quad
\Phi \left(\begin{array}{cc} \r F{} \\
\r {\overline F}{} \end{array}\right)=
\left(\begin{array}{cc}  F \\ {\overline F} \end{array}\right), \quad
{\hat\Phi} \left(\begin{array}{cc} F \\ {\overline F} \end{array}\right)=
\left(\begin{array}{cc} \r F{} \\ \r {\overline F}{} \end{array}\right),
\label{mat2}
\end{eqnarray}
where $F$ and $\overline F$ are arbitrary solutions to the symmetry
equation (\ref{5})\footnote{To check relation (\ref{mat2}) for the
first nontrivial solution (\ref{sol}), it is necessary to remove the
ambiguity in the operator ${\partial}^{-1} {\partial} 1$ that appears in
the calculations by setting ${\partial}^{-1} {\partial} 1=({\partial}^{-1}
{\partial}) 1 \equiv 1$.}. Acting by the projectors  $\Pi$ and
${\overline \Pi}$ (\ref{pi}) on the condition (\ref{inv2}), one can
immediately check that the constraint (\ref{inv2}) is consistent with the
constraint (\ref{cons}).

Let us present an infinite set of partial solutions of the condition
(\ref{inv2}).

Without going into detail, we state that the solution $J_1$ of the
condition (\ref{inv2}), with the scaling dimension $[cm^{0}]$
corresponding to the first Hamiltonian structure, has the following form:
\begin{equation}
J_1(f, {\overline f})={\phi}_1\otimes{\partial}{\phi}_1^{T},
\label{hamstr1}
\end{equation}
where ${\phi}_1$ is defined by (\ref{frech}). It is obvious that $J_1$ is
the symmetric operator satisfying the Jacobi identity. One can easily show
that $J_1$ also satisfies the chiral constraints (\ref{cons}). As
for the condition (\ref{inv2}), it can be checked by direct though
laborious calculations. However, there is an easier way to prove
(\ref{hamstr1}). Substituting $J_1$
(\ref{hamstr1}) into (\ref{inv2}) and making some obvious algebraic
transformations, the condition (\ref{inv2}) can be rewritten in the
following equivalent form:
\begin{equation}
J_1(\r {f}{}, \r {\overline f}{}) =
A {\sigma}J_1(f, {\overline f}){\sigma}^{-1} A.
\label{h1}
\end{equation}
Let us introduce the matrix $J^{*}_1$ defined by the following equation:
\begin{eqnarray}
\{ J_1 , J^{*}_1 \} = 1,
\label{propstarj}
\end{eqnarray}
which admits the unique solution. Using relations (\ref{mat2}) and their
adjoint, one can easily construct this solution$^{3}$,
\begin{eqnarray}
J^{*}_1(f, {\overline f}) = {{\hat{\phi}}_1}^{T} \otimes {\partial}^{-1}
{\hat{\phi}}_1 \equiv {\overline \Pi} \left(\begin{array}{cc} {\overline
f} \partial^{-1} {\overline f},& -{\overline f} \partial^{-1} f-2\\ -f
\partial^{-1} {\overline f}-2,& f \partial^{-1} f \end{array}\right) \Pi,
\label{starj}
\end{eqnarray}
where ${\hat{\phi}}_1$ is determined by eq. (\ref{m}).
Applying the automorphism $\sigma$ (\ref{auto}) to eqs. (\ref{propstarj})
and having in mind the one-to-one correspondence between $J_1$ and
$J^{*}_1$, one can conclude that the condition (\ref{h1}) for $J_1$
is satisfied if and only if, the similar condition for $J^{*}_1$,
\begin{equation}
J^{*}_1(\r {f}{}, \r {\overline f}{}) =
A {\sigma}J^{*}_1(f, {\overline f}){\sigma}^{-1} A,
\label{h2}
\end{equation}
is implemented. Because of the rather simple structure of $J^{*}_1$
(\ref{starj}), it is a simple exercise to check the correctness of
eq. (\ref{h2}) and, therefore, eq. (\ref{h1}) for $J_1$ is also
correct. This completes the proof of eq. (\ref{hamstr1}).

Acting $k$-times by the recursion operator (\ref{recop}) on the first
Hamiltonian structure (\ref{hamstr1}) and taking into account that the
scaling dimension $[R]=cm^{-1}$, it is easy to understand that
\begin{equation}
J_k=R~^kJ_1
\label{hamstrn}
\end{equation}
gives us the $k$-th Hamiltonian structure.

Using the general rule (\ref{hamstrn}), we derive, for example, the
following representation for the second Hamiltonian structure$^{3}$ $J_2$:
\begin{eqnarray}
J_2(f, {\overline f})=\frac{1}{2}\left(\begin{array}{cc}
  fD{\overline D} \partial^{-1} f,&
D\overline D -fD {\overline D} \partial^{-1} {\overline f}\\
-{\overline D} D +{\overline f}~{\overline D}D\partial^{-1} f, &
-{\overline f}~{\overline D}D{\partial^{-1}}{\overline f}
\end{array}\right),
\label{hamstr2}
\end{eqnarray}
which resembles the form of its bosonic counterpart and obviously satisfies
the chiral consistency conditions (\ref{cons}).

In terms of $N=1$ superfields, the explicit expressions for the first and
second Hamiltonian structures of the super-NLS hierarchy were constructed
in \cite{10}; however, to our knowledge, the recursion operator
(\ref{recop}) is presented here for the first time.

To conclude this section, we would like to stress that the consistency
conditions (\ref{cons}) is satisfied for all the Hamiltonian structures
(\ref{hamstrn}). This is evident from the explicit form of the recursion
operator $R$ (\ref{recop}) and the properties of the projectors
(\ref{pi}). Thus, all of the hamiltonian structures are degenerate
matrices. This is the peculiarity of a manifest $N=2$ superinvariant
description of the super-NLS hierarchy in terms of $N=2$ superfields,
which has no analogue in the description in terms of $N=1$ superfields or
components. This means that the standard representation of $R$ in terms
of the first and second Hamiltonian structures $R=J_2J^{-1}_1$ fails,
though the relation $RJ_1=J_2$ is correct. It is instructive to find its
correct generalization in the case under consideration. Without going into
detail, let us present the answer,
\begin{eqnarray}
R = J_2 J^{*}_1,
\label{recop1}
\end{eqnarray}
where the matrix $J^{*}_1$ is defined by eq. (\ref{starj}).
Using (adjoint) eqs. (\ref{mat2}), it is easy to verify the mutual
relations (\ref{hamstrn}) and (\ref{recop1}) between $R$, $J_1$,
$J^{*}_1$, and $J_2$.

\section{The f-Toda invariant Hamiltonians}

In this section, we construct the Hamiltonians which are invariant with
respect to the f-Toda mapping (\ref{3}) and demonstrate that the evolution
equations (\ref{evol}) can be represented in Hamiltonian form.

Let us recall that in the $N=2$ supersymmetric case, the Hamiltonian
$H(f,\overline f)$ can be expressed in terms of the Hamiltonian density
${\cal H}(f,\overline f)$ as
\begin{equation}
H(f,\overline f)=\int dZ {\cal H}(f,\overline f),
\label{meas}
\end{equation}
where $Z=(x,\theta,\overline\theta)$ is the coordinate of $N=2$ superspace
and $dZ=dxd\theta d{\overline \theta}$ is an invariant $N=2$ supersymmetric
measure.

The Hamiltonian $H(f,\overline f)$ is invariant with respect to the
f-Toda mapping (\ref{3})
\begin{eqnarray}
H(\r {f}{},\r {\overline f}{}) =  H(f,\overline f),
\label{inv3}
\end{eqnarray}
if the Hamiltonian density${\cal H}(f,\overline f)$ satisfies the following
condition:
\begin{eqnarray}
{\cal H}(\r {f}{},\r {\overline f}{}) -{\cal H}(f,\overline f)=
\Psi+ {\overline \Psi},
\label{inv1}
\end{eqnarray}
where $\Psi$ (${\overline \Psi}$) is an arbitrary local chiral
(antichiral) function of $f$ and ${\overline f}$,
\begin{equation}
D\Psi={\overline D}~{\overline \Psi}=0.
\label{ch}
\end{equation}
Let us note that the r.h.s. of the condition (\ref{inv1}) admits a more
general structure in comparison with its bosonic counterpart \cite{7,8},
where only the derivative $\partial$ ($=-\{D,\overline D\}$) of an
arbitrary local function is admitted. It is evident that through
integration over the invariant supersymmetric measure $dZ$, due to
(\ref{ch}), it becomes equal to zero, providing the invariance condition
(\ref{inv3}) for the Hamiltonian.

Now, we will construct an infinite set of partial solutions
of the condition (\ref{inv1}).

Acting by the operator $2{\partial}^{-1}$ on the symmetry equation
(\ref{5}),
\begin{equation}
{\partial}^{-1}(\r {F}{} \r {\overline f}{}+\r {f}{} \r {\overline F}{})-
{\partial}^{-1}(F \overline f+f\overline F)=
2{{\overline D} \r {F}{} \over {{\overline D} \r {f}{}}} +
2{D {\overline F} \over D {\overline f}}
\label{sym5}
\end{equation}
and comparing the result (\ref{sym5}) with (\ref{inv1}),
one can immediately find the solution to the
condition (\ref{inv1}) for the Hamiltonian density ${\cal H}$,
\begin{equation}
{\cal H}={\partial}^{-1}(F \overline f + f\overline F).
\label{dens}
\end{equation}
Substituting the infinite set of the solutions (\ref{set}) of the symmetry
equation (\ref{5}) into (\ref{dens}), one can generate the infinite set of
Hamiltonian densities ${\cal H}_p$ with scale dimension $p$
(see the paragraph below formula (\ref{evol})):
\begin{equation}
{\cal H}_p={\partial}^{-1}(F_p \overline f + f{\overline F}_p).
\label{den1}
\end{equation}
Using the explicit expressions (\ref{three}), we obtain, for example, the
following four first Hamiltonian densities:
\begin{eqnarray}
{\cal H}_1 &=& f{\overline f}, \quad
{\cal H}_2 = 2(f'{\overline f}-\frac{1}{2}(f{\overline f})'), \nonumber\\
{\cal H}_3 &=& 3(f''{\overline f} -
\frac{1}{2} f {\overline f}~{\overline D} f\cdot D {\overline f} -
(f'f)'+\frac{1}{3}(f {\overline f})''), \nonumber\\
{\cal H}_4 &=& f'''{\overline f} -f''{\overline f}'+f'{\overline f}''-
f{\overline f}'''+2 f{\overline f}f'{\overline f}' \nonumber\\
&+& f{\overline f}({\overline D} f\cdot D {\overline f}'
-{\overline D} f'\cdot D {\overline f})+
2(f {\overline f}'-f'{\overline f}){\overline D} f\cdot D {\overline f}.
\label{hami}
\end{eqnarray}
Up to unessential total derivatives and overall multipliers, these
Hamiltonian densities coincide with the corresponding quantities of the
$N=2$ super-NLS hierarchy \cite{3}, which confirms the
above-mentioned inter-relation of the f-Toda mapping (\ref{3}) and the
$N=2$ super-NLS hierarchy.

Using the explicit expressions for the f-Toda-invariant first and second
Hamiltonian structures, (\ref{hamstr1}), (\ref{starj}), and
(\ref{hamstr2}), as well as for the invariant Hamiltonians (\ref{hami}),
one can construct the Hamiltonian system of evolution equations
\begin{equation}
\frac{\partial}{\partial t_p}
\left(\begin{array}{cc} f\\{\overline f} \end{array}\right) =
J_1\left(\begin{array}{cc} {\delta}/{\delta f} \\
{\delta}/{\delta {\overline f}} \end{array}\right) H_{p+1}=
J_2\left(\begin{array}{cc} {\delta}/{\delta f} \\
{\delta}/{\delta {\overline f}} \end{array}\right) H_{p},
\label{hameq}
\end{equation}
\begin{equation}
J^{*}_1
\frac{\partial}{\partial t_p}
\left(\begin{array}{cc} f\\{\overline f} \end{array}\right) =
\left(\begin{array}{cc} {\delta}/{\delta f} \\
{\delta}/{\delta {\overline f}} \end{array}\right) H_{p+1},
\label{hameq2}
\end{equation}
which, by construction, are invariant with respect to the f-Toda mapping
(\ref{3}). Direct calculations show that they are equivalent to the
evolution equations (\ref{evol}), (\ref{three}),  i.e. the following
relation
\begin{equation}
\left(\begin{array}{cc} F_p \\ {\overline F}_p \end{array}\right) =
J_1 \left(\begin{array}{cc} {\delta}/{\delta f} \\
\delta / {\delta {\overline f}}
\end{array}\right) H_{p+1}
\label{h3}
\end{equation}
is satisfied.

We would like to close this section with the remark that the first
Hamiltonian density ${\cal H}_1$ satisfies the following equation of
motion:
\begin{equation}
\frac{\partial}{\partial t_p} {\cal H}_1 = {\cal H}_p'.
\label{hami0}
\end{equation}
Hence, there exists the additional integral of motion,
\begin{equation}
{\widetilde H}_1 = \int dx {\cal H}_1,
\end{equation}
where we have only space integration, which
means that ${\widetilde H}_1$ is the unconstrained superfield and,
therefore, it contains four independent components, the Hamiltonians. To
obtain relation (\ref{hami0}), one can substitute expressions (\ref{evol})
for $F_p$ and ${\overline F}_p$, as well as for ${\cal H}_1$ (\ref{hami})
into (\ref{den1}). At $p=2$, this property was observed in \cite{9} for
the wide class of $N=2$ supersymmetric generalized NLS hierarchies
constructed there. Taking this fact into account, as well as that $N=2$
super-NLS is a particular representative of the class of $N=2$ super-GNLS
hierarchies, it seems plausible to assume that relation (\ref{hami0})
is also satisfied for the entire class of $N=2$ super-GNLS hierarchies at
arbitrary values of the parameter $p=1,2,3, ...$, as in the $N=2$
super-NLS case.

\section{Conclusion}

In this paper, we have proposed the $N=2$ supersymmetric f-Toda mapping
(\ref{3}) in $(1|2)$ superspace that can be considered as the minimal
$N=2$ superextension of the one-dimensional Toda chain. We demonstrated
that the $N=2$ super-NLS hierarchy is invariant with respect to the f-Toda
mapping, and produced its manifestly $N=2$ supersymmetric recursion
operator and Hamiltonian structures, using only their symmetry properties.
New general representations (\ref{den1}) and (\ref{hami0}) for its
Hamiltonians were observed.

We would like to note that the f-Toda substitution is not
exotic. There is a wide class of such kinds of substitutions for which the
approach developed in the present letter may be literally
applied. We hope to present them together with the corresponding
integrable hierarchies in future publications.
It would also be interesting to construct a super-Hamiltonian structure of
the f-Toda chain and its explicit solutions.
It seems to be very important to find its two-dimensional
integrable counterparts that could admit the
superconformal structure and give new example of consistent
two-dimensional supersymmetric field theories. This work is under progress
at the present time.

\section{Acknowledgments}

This work was partially supported by the Russian
Foundation for Basic Research, Grant No. 96-02-17634, INTAS grant No.
94-2317, and by a grant from the Dutch NWO organization.

\vspace{1cm}

{\Large \bf Appendix}
\setcounter{equation}{0}
\def\theequation{A.\arabic{equation}}
\vspace{0.5cm}

Here, we shortly describe the main steps of the proof of the assertion
(\ref{7}) of section 3.

First, it is necessary to substitute $\widetilde F$ and
$\widetilde {\overline F}$ (\ref{7}) into the symmetry equation (\ref{5}).
Direct calculations give the following form for the different terms
of the symmetry equation:
\begin{eqnarray}
&& \overline D \r {\widetilde F}{}=
\overline D \r {F'}{}+
{1\over2}\overline D (\r {f}{} \r {\overline f}{}\r {F}{})-
[\r {f}{}\overline D +
((\overline D\r {f}{})D\overline D+
{1\over2}(\overline D\r {f'}{})-
{1\over 2}\r{f'}{}~\overline D)\partial^{-1}]
(\r {f}{} \r {\overline F}{}+\r {F}{}\r {\overline f}{}), \nonumber\\
&& D \widetilde {\overline F}=-D \overline F'+
{1\over 2} D (f\overline f~\overline F) -
[\overline f D+
((D\overline f)~\overline DD+{1\over2}(D\overline f')-
{1\over 2}\overline f' D)\partial^{-1}]
(f\overline F+F\overline f), \nonumber\\
&& \r {\widetilde F}{} \r {\overline f}{}+
\r {f}{} \r {\widetilde {\overline F}}{}-
{\widetilde F} {\overline f}-
f {\widetilde {\overline F}}=
\r {F'}{} \r {\overline f}{} - \r {f}{} \r {\overline F'}{}-
{1 \over 2}[(\r {f}{} \r {\overline f}{})'+
(\overline D \r {f}{}) \r {\overline f}{}D-
\r {f}{}(D \r {\overline f}{})\overline D]
\partial^{-1}(\r {f}{} \r {\overline F}{}+
\r {F}{} \r {\overline f}{}) \nonumber\\
&&~~~~~~~-F'{\overline f}+f{\overline F'}+
{1 \over 2}[(f \overline f)'+ ({\overline D} f){\overline f}D-
f(D\overline f)\overline D]\partial^{-1}(f\overline F+F\overline f).
\end{eqnarray}

Second, it is necessary to use relations (\ref{3}) and (\ref{5})
and their direct consequences: the two identities which can be obtained
  from (\ref{3}) by the action of derivatives $D$ and $\overline D$,
respectively; the identity which can be produced from (\ref{5}) by the
action of the operator $[D,{\overline D}]$; and the following identity:
\begin{equation}
({1\over 2} \r {f}{} \r {\overline f}{} -(\ln \overline D \r {f}{})')
({1\over 2}{\partial}^{-1}(\r {F}{} \r {\overline f}{}+
\r {f}{} \r {\overline F}{})-
{{\overline D} \r {F}{} \over{{\overline D} \r {f}{}}})=
({1\over 2} f{\overline f} + (\ln  D{\overline f})')
({1\over 2}{\partial}^{-1}(F{\overline f}+f {\overline F})+
{D{\overline F} \over D {{\overline f}}}),
\label{ident}
\end{equation}
which one can derive by rewriting relations (\ref{3})
and (\ref{sym5}) in the following equivalent form:
\begin{eqnarray}
{1\over 2} \r {f}{} \r {\overline f}{} -(\ln \overline D \r {f}{})'&=&
{1\over 2} f{\overline f} + (\ln  D{\overline f})', \nonumber\\
{1\over 2}{\partial}^{-1}(\r {F}{} \r {\overline f}{}+
\r {f}{} \r {\overline F}{})-
{{\overline D} \r {F}{} \over{{\overline D} \r {f}{}}}&=&
{1\over 2}{\partial}^{-1}(F{\overline f}+f {\overline F})+
{D{\overline F} \over {D{\overline f}}},
\end{eqnarray}
respectively, and equating the product on their left-hand sides to the
product on their right-hand sides.


\end{document}